\documentclass[aps,prb,showpacs,reprint, superscriptaddress]{revtex4-1}

\usepackage{graphicx}
\usepackage{color}
\usepackage{amsmath}
\usepackage{bbm}

\input epsf

\begin{document}

\title{Fulde-Ferrell state in superconducting core/shell nanowires: role of the orbital effect}
\author{Marek Mika}

\affiliation{AGH University of Science and Technology, Faculty of
Physics and Applied Computer Science, 30-059 Krakow, Poland
Al. Mickiewicza 30, 30-059 Krakow, Poland}
\author{Pawe{\l} W{\'o}jcik}
\email{pawel.wojcik@fis.agh.edu.pl}
\affiliation{AGH University of Science and Technology, Faculty of
Physics and Applied Computer Science, 30-059 Krakow, Poland
Al. Mickiewicza 30, 30-059 Krakow, Poland}

\date{03.02.2015}

\begin{abstract}
The orbital effect on the Fulde-Ferrell (FF) phase is investigated in superconducting core/shell nanowires subjected to the axial magnetic
field. The confinement in the radial direction results in the quantization of the electron motion with energies determined by the radial $j$ and
orbital $m$ quantum numbers. In the external magnetic field the twofold degeneracy with respect to the orbital magnetic quantum number $m$ is lifted
which leads to the Fermi wave vector mismatch between the paired electrons $(k, j,m,\uparrow) \leftrightarrow (-k, j,-m,\downarrow)$. This mismatch
is transfered to the nonzero total momentum of the Cooper pairs which results in the formation of FF phase occurring sequentially with increasing magnetic field. 
By changing the nanowire radius $R$ and the superconducting shell thickness $d$, we discuss the role of the orbital effect in the
FF phase formation in both the nanowire-like ($R/d \ll 1$) and nanofilm-like ($R/d \gg 1$) regime. We have found that the irregular pattern
of the FF phase, which appears for the case of the nanowire-like regime, evolves towards the regular distribution, in which the FF phase stability regions 
appear periodically between the BCS state, for the nanofilm-like geometry.  The crossover between these two different phase diagrams is explained as resulting from the orbital
effect and the multigap character of superconductivity in core/shell nanowires.
\end{abstract}


\maketitle
\section{Introduction}
In the last decade, unconventional superconductivity with a nontrivial Cooper pairing 
has attracted the growing interest due to fascinating superconducting properties which are 
not observed for the standard BCS state. Among the wide class of unconventional superconductors including 
high-$T_c$ cuprates,\cite{Tabis2014,Badoux2016,Zegrodnik2017} iron-pnictides,\cite{Paglione2010} or heavy fermion
materials\cite{Matsuda2007}  recently, the special attention is  drawn to systems with a 
spatially varying energy gap.\cite{Hamidian2017}
The existence of such a superconducting phase with the order parameter oscillating in real space
was proposed in the mid-1960s by Fulde and Ferrell\cite{Fulde1964} (FF phase) as well as independently by Larkin and 
Ovchinnikov (LO phase).~\cite{Larkin1964} According to their original concept, superconductivity can survive 
in the magnetic field substantially higher than the critical field $H_{c}$, due to the creation 
of an inhomogeneous paired state with a non-zero total momentum of the Cooper pairs 
$(\mathbf{k}\uparrow,-\mathbf{k}+\mathbf{q} \downarrow)$. This so called FFLO state results
from the pairing between electrons from Zeeman splitted parts of the Fermi surface.

In spite of a straightforward nature of the theoretical prediction and many ongoing theoretical investigations
regarding the appearance of the FFLO state in different materials,\cite{Kaczmarczyk2010,Maska2010,Ptok2013_1,Ptok2014_1,Zegrodnik2014} the
experimental evidence of the non-zero momentum pairing has been reported only recently in heavy fermion 
systems\cite{Matsuda2007,Blanchi2003,Kumagai2006,Correa2007} and two dimensional organic
superconductors.\cite{Beyer2013,Singleton2000,Tanatar2002,Shinagawa2007} Both of these material classes are characterized by a 
reduction of the orbital pair breaking mechanism which is a crucial physical limitation for the experimental realization of the FFLO 
phase. The significance of the orbital pair braking is described by the Maki parameter\cite{Maki1966,Maki1964} 
defined as  $\alpha=\sqrt{2} H_{c2}^{orb}/H_{c2}^{P}$, where $H_{c2}^{orb}$ is the upper critical field 
calculated without Zeeman splitting and $H_{c2}^{P}$ is the critical field in paramagnetic limit.\cite{Clogston1962,Chandrasekhar1962} 
It has been established that the FFLO phase can exist at finite temperature if $\alpha>1.8$.\cite{Gruenberg1966} 
This criterion can be met in 
ultrathin metallic nanofilms in which the confinement in the direction perpendicular 
to the film strongly reduces the orbital effect for the in-plane magnetic field. The
theoretical model describing the FFLO phase in metallic nanofilms, besides the possibility of the non-zero 
momentum pairing, should also contain the multiband character of supercondcutivity in these systems.
In metallic nanostructures with size comparable to the electron wave length, the Fermi surface splits into a set of discrete subbands leading
to many interesting effects which are not observed in the bulk limit e.g, the formation of Andreev
states\cite{Shanenko_andreev} or oscillations of superconducting
properties.\cite{Zhang2010,Uchihashi2011,Qin2009,Zhang2010,Ozer2006_b,Ozer2007,Guo2004,Zhang2010,Wojcik2014_1,Wojcik2014_3,Shanenko2010,Shanenko2008}
As reported in our recent paper,\cite{Wojcik2016} due to the multiband nature, the Fulde-Ferrell (FF) phase in metallic nanofilms splits into 
subphases, number of which corresponds to the number of subbands participating in the formation of the 
paired state. Similar behavior has been also reported for a Pauli-limiting two-band superconductors.~\cite{Takahashi2014} 
In both of these reports the FF phase has been induced by the Zeeman effect for the magnetic field $H>H_c$.

The multiband character of superconductivity is even more pronounced in metallic nanowires. Very interesting 
phenomenon has been recently found when studying the superconducting to normal metal transition in nanowires, driven 
by the axial magnetic field.\cite{Shanenko2008} It turned out that the magnetic field does not destroy superconductivity simultaneously 
in all subbands participating in the paired phase but the transition to the normal state occurs gradually. 
The magnetic field suppress superconducting correlations step by step in subsequent subbands. It reveals itself as a cascade of jumps in the order 
parameter with increasing magnetic field. Such anomalous behavior has inspired our recent study\cite{Wojcik2015} in which, 
surprisingly, we have found that in cylindrical nanowires subjected to the axial magnetic field, the orbital effect, which 
so far has been regarded as detrimental to the FFLO phase formation, can in fact induce the non-zero momentum paired state. 
As shown in Ref.~\onlinecite{Wojcik2015}, the Fermi wave vector mismatch induced by the orbital effect between the subbands with opposite orbital
momenta is transfered to the nonzero total momentum of the Cooper pairs which results in the formation of sequentially occurring Fulde-Ferrell (FF)
and BCS phases with increasing magnetic field. In this context, understanding the physical mechanism standing behind the change of phase diagrams from the Pauli-limit, in which FF
phase occurs in the vicinity of $H_c$ as for nanofilms, to the orbital limit, in which the stable FF phases appear between BCS-paired states for
$H<H_c$, still remains an unexplored issue. This can be done by considering superconducting core/shell nanowires, in
which, by the control of the ratio $R/d$, where $R$ is the core radius and $d$ is the shell thickness, we can switch from the
nanowire-like ($R/d \ll 1$) to the nanofilm-like ($R/d \gg 1$) scenario.\cite{Chen2012,Chen2010}

In the present paper, by controlling the ratio $R/d$,  we discuss the role of the orbital effect in the FF phase formation, in both the
nanowire-like and nanofilm-like regime. We have found that the phase diagrams differ considerably in both of these regimes. The irregular pattern of
the FF phase occurrence in the nanowire-like regime evolves, with increasing the $R/d$ ratio, towards the regular one, in which the FF phases appear
periodically between the BCS states. The crossover between these two different phase diagrams is explained as resulting
from the orbital effect and the multigap character of superconductivity in the considered nanostructures.

The paper is organized as follows. In the next section we introduce the basic concepts of the theoretical model based on the modified BCS theory, in
which the superconducting gap acquires the non-zero total momentum of the Cooper pairs. We explain in detail how the angular-momentum-induced
Fermi-surface splitting generates the Fulde-Ferrell phase. In Sec.~\ref{sec:results} we discuss our results considering the contributions of both the
orbital and Zeeman effect to the FF state. Finally, Sec.~\ref{sec:conclusion} is devoted to conclusions and short discussion on the possibility of
the experimental verification of the phenomena presented in the paper. 

\section{Theoretical model}
\label{sec:theory}
\begin{figure}[ht]
\begin{center}
\includegraphics[scale=0.23]{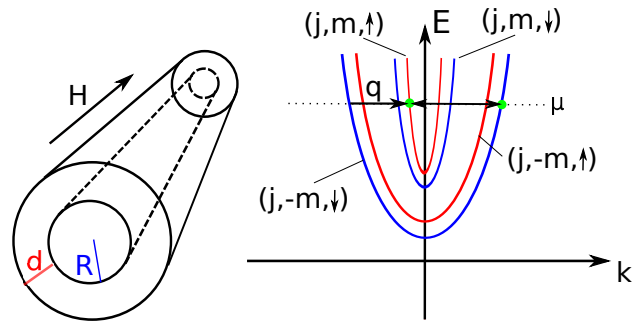}
\caption{(Color online) (a) Scheme of the superconducting core/shell nanowire. (b) Schematic illustration of FF-pairing in the nanowire.  
In the presence of the magnetic field a two-fold degeneracy with respect to the orbital magnetic quantum number $m$ and two-fold degeneracy with
respect to the spin $\sigma$ are lifted. The Fermi vector mismatch is compensated by the non-zero center-of-mass momentum of the Cooper pairs.}
\label{fig1}
\end{center}
\end{figure}

Let us consider the core/shell nanowire consisting of a core of radius $R$, surrounded by a superconducting shell of thickness $d$
[Fig.~\ref{fig1}(a)]. 
Recently, analogous systems of semiconductor nanowires covered by a superconducting layer, have attracted 
growing interest due to their potential application in topologically protected quantum computing using Majorana zero
modes.\cite{Mourik2012,Krogstrup2015}  For simplicity, let us assume that the core is an ideal insulator and electrons cannot
penetrate the region of the core which allows us to neglect the proximity effect at the superconductor/insulator interface. We start from the
general form of the BCS Hamiltonian 
\begin{equation}
 \begin{split}
 \hat{\mathcal{H}}&=\sum _{\sigma} \int d^3 r
\hat{\Psi}^{\dagger} (\mathbf{r},\sigma) \hat{H}_0
\hat{\Psi}(\mathbf{r},\sigma) \\ 
&+ \int d^3 r \left [ \Delta
(\mathbf{r})\hat{\Psi}^{\dagger}(\mathbf{r},\uparrow)
\hat{\Psi}^{\dagger}(\mathbf{r},\downarrow) + h.c. \right ]\\
& +\int d^3r \frac{|\Delta(\mathbf{r})|^2}{g},
 \end{split}
 \label{eq:BCS_general}
\end{equation}
where $\sigma$ denotes the spin state $(\uparrow,\downarrow)$, $g$ is the phonon-mediated electron-electron coupling constant
and the gap parameter in real space is given by
\begin{equation}
 \Delta(\mathbf{r})=-g \left < \hat{\Psi} (\mathbf{r},\downarrow)
\hat{\Psi} (\mathbf{r},\uparrow)  \right >.
\label{eq:gap_def}
\end{equation}
Choosing the gauge for the vector potential as $\mathbf{A}=(0,eHr / 2,0 )$, where the magnetic field $H$ is directed along the nanowire
axis, the single-electron Hamiltonian $\hat{H}_0$ in the cylindrical coordinates $(r,\varphi,z)$ is given by
\begin{equation}
 \hat{H}_0=\frac{\hbar ^2}{2m_e} \left [ -\frac{1}{r}
\frac{\partial}{\partial r} r \frac{\partial}{\partial r} + \left ( -
\frac{i}{r} \frac{\partial}{\partial \varphi} + \frac{eHr}{2\hbar}
\right ) ^2 - \frac{\partial ^2}{\partial z ^2} \right ] + \sigma \mu _B H - \mu \;,
\label{ham1D}
\end{equation}
where $\sigma=\pm 1$ for spin-up and spin-down electrons, $\mu$ is the chemical potential and $e$, $m_e$ is the electron charge and mass, 
respectively. \\
If we assume azimuthal invariance and neglect the diamagnetic term $\sim\mathbf{A}^2$, whose energy for nanowires 
is one order of magnitude lower than the order parameter, $\hat{H}_0$ can be reduced to the one-dimensional form 
\begin{equation}
 \hat{H}_{0,1D}=\frac{\hbar ^2}{2m_e} \left [ -\frac{1}{r}
\frac{\partial}{\partial r} r \frac{\partial}{\partial r} + \frac{m^2}{r^2} \right ] + \frac{\hbar^2 k^2 }{2m_e} + (m+\sigma) \mu _B H
 - \mu \;,
\label{eq:H1dd}
\end{equation}
with the corresponding single-electron wave functions
\begin{equation}
 \psi _{k,j,m}(r,\varphi,z)=\frac{1}{\sqrt{2 \pi L}} \phi _{j,m}(r)e^{im\varphi}e^{ikz},
\end{equation}
where $L$ is the nanowire length, $j$ is the radial quantum number, $m$ is the orbital magnetic quantum number and 
$k$ is the wave vector along the nanowire axis $z$. By assuming the hard-wall boundary conditions in the shell,
$\phi _{jm}(R)=\phi _{jm}(R+d)=0$, the radial wave function $\phi _{jm}(r)$ can be written as\cite{Chen2010}
\begin{equation}
 \phi _{j,m}(r)=\frac{1}{\sqrt{\mathcal{M}}} \left [ Y_m(\chi _{jm} R) J_m(\chi _{jm} r) -  J_m(\chi _{jm} R) Y_m(\chi _{jm} r) \right ],
\end{equation}
where $J_m(r)$ and $Y_m(r)$ are the Bessel functions of the first and second kind of $m$-th order and $\mathcal{M}$ is the normalization constant. 
The parameter $\chi _{jm}$, related to the single-electron energy $\xi _{k,j,m,\sigma}$ by 
\begin{equation}
\xi _{k,j,m,\sigma}=\frac{\hbar ^2}{2m_e} ( \chi _{jm}^2 + k^2 ) + (m+\sigma) \mu _B H - \mu,
\label{eq:single_energy}
\end{equation}
is a solution of the equation
\begin{equation}
Y_m(\chi _{jm} R) J_m[\chi _{jm} (R+d)] -  J_m[\chi _{jm} (R+d)] Y_m(\chi _{jm} R)=0. 
\end{equation}
From Eq.~(\ref{eq:single_energy}) we can see that for $H=0$, each single-electron state is fourfold degenerate - two-fold degeneracy with respect
to the orbital magnetic quantum number $m$ and two-fold degeneracy with respect to the spin $\sigma$. In the presence of external magnetic field
both these degeneracies are lifted resulting in a shift between the subbands corresponding to $m$ and $-m$ as well as $\uparrow$ and $\downarrow$.
Since in the superconducting state the pairing appears between particles with opposite
spins, momenta and orbital momenta: $(k, j,m,\uparrow) \leftrightarrow (-k, j,-m,\downarrow)$, the Fermi-wave vector mismatch induced in the magnetic
field can be transferred into the non-zero momentum of the Cooper pairs ($q \ne 0$ along the $z$ axis) giving raise to the FF phase. Schematic
illustration of this process is sketched in Fig.~\ref{fig1}(b).  Using the field operators in the form 
\begin{equation}
\begin{split}
 \hat{\Psi}(r,\varphi,z,\sigma)=\sum_{k,j,m}
\psi_{k,j,m}(r,\varphi,z)\:\hat{c}_{k,j,m,\sigma},\\
\hat{\Psi}^{\dagger}(r,\varphi,z,\sigma)=\sum_{k,j,m}
\psi^*_{k,j,m}(r,\varphi,z)  \:
\hat{c}^{\dagger}_{k,j,m,\sigma},
\end{split}
\label{eq:field_op}
\end{equation}
where $\hat{c}_{k,j,m,\sigma} (\hat{c}^{\dagger}_{k,j,m,\sigma})$ is the annihilation (creation) operator, the BCS Hamiltonian with the possibility
of non-zero momentum pairing is given by 
\begin{equation}
\begin{split}
\hat{H}&=
\sum_{kmj}\mathbf{\hat{f}}^{\dagger}_{k,j,m,q}\mathbf{H}_{k,j,m,q}\mathbf{\hat{f}}_{k,j,m,q}+\sum_{k,j,m}\xi_{-k+q,j,-m,\bar {\sigma}} \\
&+\sum_{j,m}
\frac{|\Delta_{j,m,q}|^2}{g},  
\end{split}
\label{eq:Ham_matrix_1}
\end{equation}
where
$\mathbf{\hat{f}}^{\dagger}_{k,j,m,q}=(\hat{c}^{\dagger}_{k,j,m,\uparrow}, \hat{c}_{-k+q,j,-m,\downarrow})$ is the composite vector operators and
\begin{equation}
\mathbf{H}_{k,j,m,q}=\left(\begin{array}{cc}
\xi_{k,j,m,\sigma}& \Delta_{j,m,q} \\
\Delta^*_{j,m,q} & -\xi_{-k+q,j,-m,\bar {\sigma}} \\
\end{array} \right).
\label{eq:matrix_H}
\end{equation}
In the above, for simplicity, we limit to the situation in which all the Cooper pairs have a single momentum $q$. This assumption corresponds to the
Fulde-Ferrel phase. In Eq.~(\ref{eq:matrix_H}), $\Delta_{j,m,q}$ is the superconducting energy gap in the subband $(j,m)$ defined as
\begin{equation}
\Delta _{j,m,q}=\frac{g}{4 \pi^2} \sum _{k,j',m'} C_{j,m,j',m'}
\langle\hat{c}_{-k+q,j, -m,\downarrow}\hat{c}_{k,j,m,\uparrow}\rangle
\label{eq:delta_def}\;,
\end{equation}
with the interaction matrix 
\begin{equation}
C_{j,m,j',m'} = \int_{R}^{R+d} dr \: r \: \phi _{j,m}^2(r) \phi_{j',m'}^2(r)\;.
\end{equation}
Hamiltonian (\ref{eq:Ham_matrix_1}) can be diagonalized by the Bogoliubov-de Gennes transformation 
\begin{equation}
\left(\begin{array}{cl}
        \hat{c}_{k,j,m,\uparrow}\\
        \hat{c}^{\dagger}_{-k+q,j,-m,\downarrow}\\
       \end{array}\right)=\left(\begin{array}{cc}
U_{k,j,m,q} & V_{k,j,m,q}  \\
-V_{k,j,m,q} & U_{k,j,m,q} \\
\end{array}\right)\left(\begin{array}{cl}
        \hat{\alpha}_{k,j,m,q}\\
        \hat{\beta}^{\dagger}_{k,j,m,q}\\
       \end{array}\right),
\label{eq:Bogolobov_trans}
\end{equation}
where
\begin{equation}
 \begin{split}
  U^2_{k,j,m,q}=\frac{1}{2}\bigg(1+\frac{\xi_{k,j,m,\sigma}+\xi_{-k+q,j,-m,\bar{\sigma}}}{\sqrt{(\xi_{k,j,m,\sigma}+\xi_{-k+q,j,-m,\bar{\sigma}})^2+4\Delta_{j,m,q}^2}}\bigg),\\
  V^2_{k,j,m,q}=\frac{1}{2}\bigg(1-\frac{\xi_{k,j,m,\sigma}+\xi_{-k+q,j,-m,\bar{\sigma}}}{\sqrt{(\xi_{k,j,m,\sigma}+\xi_{-k+q,j,-m,\bar{\sigma}})^2+4\Delta_{j,m,q}^2}}\bigg),
 \end{split}
\end{equation}
are the Bogoliubov coherence factors.
As a result, one obtains the following form of the quasiparticle energies
\begin{equation}
\begin{split}
  E^{\pm}_{k,j,m,q}&=\frac{1}{2}( \xi _{k,j,m,\sigma}-\xi _{-k+q,j,-m,\bar{\sigma}})  \\
  &\pm \sqrt{\frac{1}{4}( \xi _{k,j,m,\sigma} + \xi _{-k+q,j,-m,\bar{\sigma}})^2 +\Delta_{j,m,q} ^2} \\ 
  & +(m+\sigma)\mu_B H.
\end{split}
\label{eq:quasi_dis_rel}
\end{equation}
By substituting Eq.~(\ref{eq:Bogolobov_trans}) into Eq.~(\ref{eq:delta_def}) we derive the self-consistent equations for the superconducting gaps
\begin{eqnarray}
 \Delta _{j',m',q}&=&\frac{g}{4 \pi^2} \int dk \sum _{j,m}
C_{j,m,j',m'} \nonumber \\ 
&\times& \frac{\Delta _{j,m,q} \left [ 1- f(E^{+}_{k,j,m,q}) -
f(E^{-}_{k,j,m,q}) \right ]}{ \sqrt{( \xi _{k,j,m,\sigma} + \xi _{-k+q,j,-m,\bar{\sigma}})^2 +4
\Delta_{j,m,q} ^2} },
\label{eq:delta_self}
\end{eqnarray}
where $f(E)$ is the Fermi-Dirac distribution. The summation in Eq.(\ref{eq:delta_self}) is carried out only over the single-energy states
$\xi _{k,j,m,\sigma}$ inside the Debye widow $|\xi _{k,j,m,\sigma}|<\hbar \omega _D$, where $\omega _D$ is the Debye frequency. 
Since the chemical potential in nanostructures strongly deviates from that assumed in the bulk, for each shell thickness we determine $\mu$
keeping a constant electron concentration
 \begin{eqnarray}
 n_e&=&\frac{1}{\pi^2[(R+d)^2-R^2]} \int dk \sideset{}{'}\sum_{j,m}  \int^{R+d}_{R} dr \: r \nonumber \\
 &\times& \big \{ |U_{k,j,m,q} \phi _{jm}(r)|^2 f(E^+_{k,j,m,q}) \nonumber \\  
&& + |V_{k,j,m,q}\phi _{jm}(r)|^2[1-f(E^-_{kmjq})]\big \} .
\label{eq:mu_self}
\end{eqnarray}
In the considered nanowires, the spatial dependence of the superconducting gap results not only from the creation of the FF phase
$[\Delta(r,\varphi,z)=\Delta(r,\varphi)e^{iqz}]$ but it is also induced by the quantum confinement. The spatial dependence of the order
parameter in the radial direction can be expressed as
\begin{eqnarray}
 \Delta_q(r)&=&\frac{g}{4\pi ^2} \int
dk \sideset{}{'}\sum_{jm} |\phi _{jm}(r)|^2 \nonumber \\
 &\times&\frac{\Delta _{j,m,q}}{\sqrt{( \xi _{k,j,m,\sigma} + \xi _{-k+q,j,-m,\bar{\sigma}})^2 +4
\Delta_{j,m,q} ^2} } \nonumber \\
&\times& \left [ 1- f(E^{+}_{k,j,m,q}) - f(E^{-}_{k,j,m,q}) \right ].
 \label{eq:delta_space}
\end{eqnarray}
To obtain the phase diagram, the superconducting gaps $\Delta_{j,m,q}$ and the chemical potential are
calculated by solving Eqs. (\ref{eq:delta_self}) and (\ref{eq:mu_self}) self-consistently. The wave-vector $q$ is determined
by minimizing the free energy of the system.\cite{Kosztin1998}

Calculations presented in the paper have been carried out for the material parameters typical of aluminum:
$\hbar \omega_D=32.31$~meV, $gN(0)$=0.18, where $N(0)=mk_F / 2\pi^2 \hbar ^2$ is the bulk density of states at
the Fermi level, $\Delta _{bulk}=0.25$~meV and the chemical potential $\mu _{bulk}=0.9$~eV which corresponds to the 
electron density $n_e=3.88 \times 10^{21}$~cm$^{-3}$. The assumed low value of the chemical potential, in relative to that
measured in the bulk, results from the parabolic band approximation (for more details, see Ref.~\onlinecite{Shanenko2006}).
Its value has been determined to obtain a good agreement with the experimental data reported in 
Ref.~\onlinecite{Shanenko2006_exp}. The self-consistent procedure has been carried out for a constant electron 
concentration which implies a gradual increase of the chemical potential with decreasing shell thickness. 
Moreover, we do not include a thickness-dependent change in the electron-phonon coupling,\cite{Hwang2000} as it can only result in the quantitative 
effects and do not alter the qualitative picture of the FF phase creation presented in the paper.

\section{Results and discussion}
\label{sec:results}
\begin{figure}[ht]
\begin{center}
\includegraphics[scale=1.0]{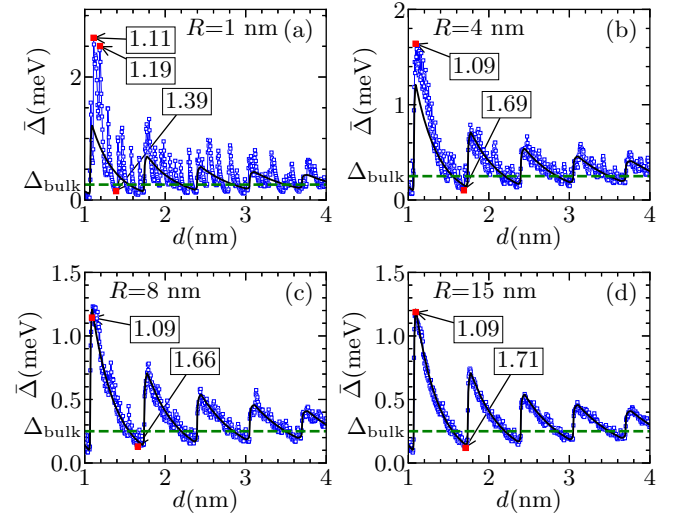}
\caption{(Color online) Spatially averaged superconducting order parameter $\bar {\Delta}$ as a function of the the shell thickness for core radii (a)
$R=1$~nm, (b) $R=4$~nm, (c) $R=8$~nm and (d) $R=15$~nm. For comparison, in each panel, $\bar {\Delta}(d)$ for a superconducting nanofilm, where $d$
is the nanofilm thickness, is shown by the solid black curve. Red, labeled points denote thicknesses chosen for the further analysis of the
unconventional FF pairing with the nonzero total momentum. Note that the same results were obtained in Ref.~\onlinecite{Chen2010} (cf. Fig.~2).}
\label{fig2}
\end{center}
\end{figure}

To determine geometrical parameters appropriate for the analysis of the non-zero total momentum pairing,
we have calculated the spatially averaged superconducting order parameter $\bar {\Delta}$, defined as
\begin{equation}
 \bar {\Delta} = \frac{2}{d(2R+d)}\int _R ^{R+d} dr \: r \Delta (r),
\end{equation}
as a function of the shell thickness for different core radii (Fig.~\ref{fig2}).

The $\bar {\Delta}(d)$ oscillations presented in Fig.~\ref{fig2} are due to the quantum size effect which arises when the system size
becomes comparable to the electron Fermi wave length.\cite{Shanenko2006} In core/shell nanowires, a reduction of an electron motion in the radial
direction implies the energy quantization with energies determined by the quantum numbers $j$, $m$, $k$ and
$\sigma$ [Eq.~(\ref{eq:single_energy})]. Subsequent peaks in $\bar {\Delta}(d)$ correspond to subsequent subbands $(j,m)$ passing
through the Fermi level while increasing the shell thickness. As seen, the $\bar {\Delta}(d)$ oscillations presented in Fig.~\ref{fig2}(a-d) differ
significantly from each other. The irregular oscillations for $R=1$~nm [Fig.~\ref{fig2}(a)], reminiscent of these predicted
for superconducting nanowires,\cite{Shanenko2006} evolves with increasing $R$ towards the regular oscillations characteristic for superconducting
nanofilms [Fig.~\ref{fig2}(d)].\cite{Shanenko2007} 
The crossover from an irregular pattern to the regular regime was explained in details in Ref.~\onlinecite{Shanenko2006}.
It is related to the centrifugal term, $\hbar ^2 m^2 / 2m_e r^2$, which for $R /d \le 1$ contributes significantly to the single electron
energy leading to the energetically well separated states for different $|m|$. The irregular oscillations of $\bar{\Delta}(d)$ presented in
Fig.~\ref{fig2}(a) reflect the irregular distribution of states $(j,m)$ on the energy scale. For $R/d \gg 1$ [Fig.~\ref{fig2}(d)] the centrifugal term
is negligibly small which causes the single electron states with different $|m|$ to be almost degenerate. These energetically close
subbands create bands labeled by the radial quantum number $j$. Each time when the bottom of such a band passes through the Fermi level we observe a
resonant increase in $\bar {\Delta}(d)$. Equal distant between bands on the energy scale results in regular oscillations presented in
Fig.~\ref{fig2}(d). Therefore, by an appropriate choice of the geometrical parameters, we can strengthen or suppress the
centrifugal energy term, which allows for a smooth transition from the nanowire-like to the nanofilm-like regime. 

\begin{figure}[ht]
\begin{center}
\includegraphics[scale=1.0]{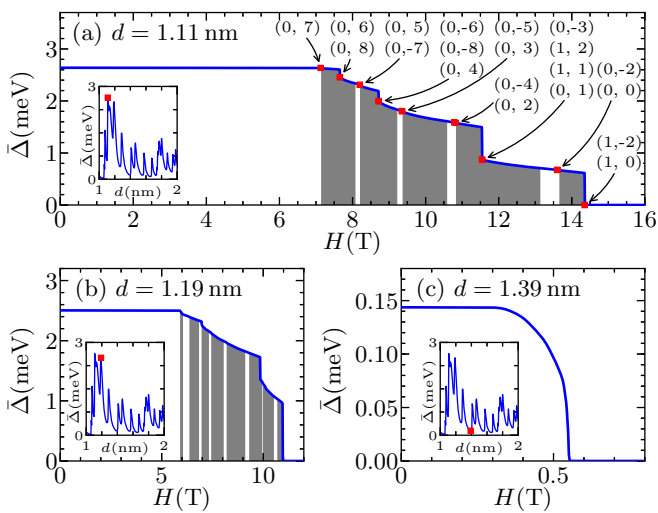}
\caption{(Color online) Magnetic field dependence of the averaged superconducting gap $\bar {\Delta}(H)$ for shell thicknesses $d$ denoted by red
squares in Fig.~\ref{fig2}(a) (see also insets). Gray areas represent the FF phase stability regions between which the conventional BCS phase, displayed as white
regions, is stable. In panel (a), the values of the magnetic field corresponding to depairing in subsequent subbands $(j,m)$ are marked by
arrows.}
\label{fig3}
\end{center}
\end{figure}

Now, let us analyze in detail the contribution of the orbital effect, $m\mu _B H$, to the FF paired phase in both of the considered regimes. We start
our study from the nanowires with $R=1$~nm. In Fig.~\ref{fig3}, the magnetic
field dependence of the averaged superconducting order parameter is presented for values of $d$ marked by red squares in Fig.~\ref{fig2}(a) which
correspond to the resonant ($d=1.11$~nm and $d=1.19$~nm) and off-resonant ($d=1.39$~nm) thicknesses, respectively.
\begin{figure}[ht]
\begin{center}
\includegraphics[scale=1.0]{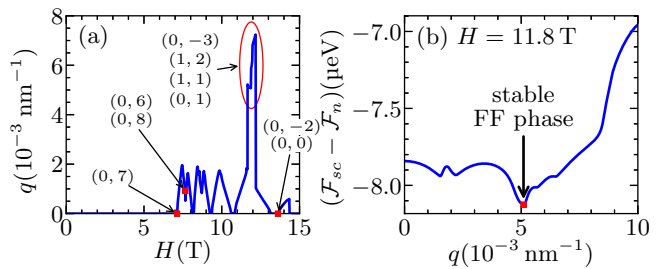}
\caption{(Color online) (a) Total Cooper-pair momentum $q$ which minimizes the free energy as a function of magnetic field $H$ together with 
(b) an exemplary difference between free energy in the superconducting and normal states $(\mathcal{F}_{sc}-\mathcal{F}_n)(q)$. Minimum of 
$(\mathcal{F}_{sc}-\mathcal{F}_n) (q)$ corresponds to the stable FF phase.}
\label{fig4}
\end{center}
\end{figure}
While increasing the magnetic field, electrons in different states $(j,m)$ acquire different energies which depend on the orbital magnetic quantum
number $m$ and the spin $\sigma$ [Eq.~(\ref{eq:H1dd})].  As a result, the superconductor to normal metal transition occurs as a cascade of
jumps\cite{Shanenko2008} (Fig.~\ref{fig3}), each of which is related to depairing in one of the subbands contributing to the superconducting state.
In Fig.~\ref{fig3}(a), the critical fields $H_c^{j,m}=\Delta _{jm}(H=0)/(|m+1|)\mu_B$ for particular subbands $(j,m)$ are labeled and marked by arrows.
Each time the magnetic field $H$ becomes slightly larger than $H_c^{j,m}$, to sustain superconductivity in the subband $(j,m)$, the Fermi
wave-vector mismatch between the paired electrons $(k,j,m,\uparrow)\leftrightarrow (-k,j,-m,\downarrow)$ is partially compensated by the non-zero
total momentum of the Cooper pairs $(k,j,m,\uparrow)\leftrightarrow (-k+q,j,-m,\downarrow)$. The formation of the FF phase minimizes the free energy
of the system, as shown in Fig.~\ref{fig4}(b). The further increase of $H$, well above $H_c^{j,m}$, causes that the paired state with the non-zero
total momentum becomes energetically less favorable and the standard BCS pairing is restored. This leads to the phase diagram  in which the FF
phase stability regimes alternate with the BCS state [Fig.~\ref{fig3}(a,b)]. 

\begin{figure}[ht]
\begin{center}
\includegraphics[scale=1.0]{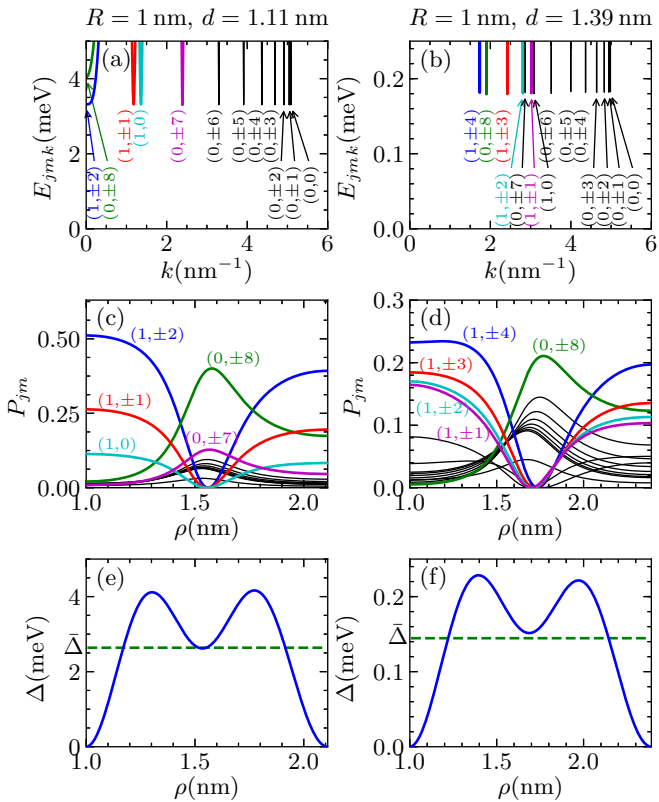}
\caption{(Color online) (a,b) Quasiparticle dispersions $E_{kjm}$ vs $k$, (c,d) contributions to the paired phase coming from
different subbands $P_{j,m}$ and (e,f) the position dependent superconducting order parameter $\Delta (r)$. Results for the resonant shell thickness
$d=1.11$~nm (left panels) and the non-resonant shell thickness $d=1.39$~nm (right panels), for $H=0$. In panels (a-d) lines plotted in color
correspond to
the subbands with the major contribution to the superconducting state.}
\label{fig5}
\end{center}
\end{figure}

Now, we discuss in detail the phase diagram for $d=1.11$~nm presented in Fig.~\ref{fig3}(a). For the chosen resonant thickness, the full spectrum of
$q$ in the whole range of the magnetic field is plotted in Fig.~\ref{fig4}(a). For completeness, in Fig.~\ref{fig5}(a) and (e) we present the
quasiparticle dispersions $E_{kjm}$ vs $k$ and the superconducting order parameter $\Delta (r)$ calculated for $H=0$. As one can see, there are twenty two
relevant subbands participating in the superconducting state: $(0,0) - (0,\pm 8)$ and $(1,0)-(1,\pm 2)$. Their contributions $P_{j,m}(r)$ to
$\Delta (r,H=0)$ are displayed in Fig.~\ref{fig5}(c), where
\begin{eqnarray}
 P_{j,m}(r)&=&\frac{g}{4\pi ^2} \int 
dk   |\phi _{jm}(r)|^2 \nonumber \\
 &\times&\frac{\Delta _{j,m,q}}{\sqrt{( \xi _{k,j,m,\sigma} + \xi _{-k+q,j,-m,\bar{\sigma}})^2 +4
\Delta_{j,m,q} ^2} } \nonumber \\
&\times& \left [ 1- f(E^{+}_{k,j,m,q}) - f(E^{-}_{k,j,m,q}) \right ].
\end{eqnarray}
Note, that the states $(1,-1)$ and $(0,-1)$, even though they contribute to the superconductivity [see Fig.~\ref{fig5}(a)], are not labeled in
Fig.~\ref{fig3}(a) as the Cooper pairs $(k,j,-1,\uparrow)\leftrightarrow (-k,j,1,\downarrow)$ are unaffected by the magnetic field i.e., for both the
electrons from the Cooper pair $(m+\sigma)\mu_B H=0$. Although, their critical fields $H^{j,-1}_c$ seem to be infinite, in
fact, the magnetic field causes the Cooper pair breaking in these states indirectly, by the reduction of the superconducting correlation in 
subbands with $m \ne -1$ [see Eq.~(\ref{eq:delta_self})]. Finally, the states $(1,-1)$ and $(0,-1)$ become depaired in $H_c$
determined for the whole nanowire. 
As presented in Fig.~\ref{fig5}(c), the contributions of the individual subbands to the superconducting order parameter, $P_{j,m}(r)$, vary
significantly. Due to the enhanced density of states, they are the largest for subbands situated in the vicinity of the Fermi surface. For
$d=1.11$~nm, the major contribution to $\Delta(r)$ comes from the states $(0,\pm 8)$ and $(1, \pm 2)$. The rest of the subbands play less important role but
the most significant contribution is due to the states $(0,\pm 7)$, $(1, 0)$ and $(1, \pm 1)$. With increasing magnetic field the
superconducting correlations are suppressed successively in individual subbands. For the subband $(j,m)$, the critical field $H_c^{j,m}$, in which
the superconductivity is destroyed, depends not only on the orbital magnetic quantum number $m$ but also on the energy gap of excitation $\Delta
_{j,m}$ [Eq.~(\ref{eq:delta_self})]. The latter is considerably affected by the quantum confinement and the Andreev mechanism, which appears due to
the spatial variation of the superconducting order parameter.\cite{Chen2010} One should note that, in nanowire-like regime, $\Delta _{j,m}$ may be
different for different quantum numbers leading to the multigap superconductivity. Therefore, the condition $H_c^{j,m_1}<H_c^{j,m_2}$ for
$|m_1|>|m_2|$ do not have to be satisfied. This expectation agrees with our numerical results showing that $H _c ^{0,8} > H _c^{0,7}$ [see
Fig.~\ref{fig3}(a)]. Consequently, the subband $(0,7)$ is the first one in which the superconducting phase is destroyed as the magnetic
field increases. The Cooper pair breaking in this single branch entails the formation of the FF phase with the total
momentum $q$ which increases with increasing magnetic field [Fig.~\ref{fig4}(a)]. This FF phase region, shown in Fig.~\ref{fig3}(a) by the gray area,
is stable up to the magnetic field value at which the Cooper pair breaking takes place in the next two states $(0,6)$ and $(0,8)$. Their
critical magnetic fields $H_c^{0, 6}$ and $H_c ^{0, 8}$ are almost equal leading to the substantial jump in $\bar{\Delta}(H)$. Preservation of
superconductivity in these branches requires to adjust a new value of the Cooper pair momentum $q$ which is shown as a sharp dip in $q(H)$
[Fig.~\ref{fig4}(a)], after which $q$ starts to increase again up to the magnetic field value at which the ordinary BCS phase is restored. Note that
in the presence of magnetic field the Fermi vector mismatch for each of the subbands $(j,m)$ is different which means that each of them has its own
favorable total momentum $q_{j,m}$. However, the situation where several values of $q_{j,m}$ appear in the system is impossible due to the coupling
between all the branches participating in the superconducting state (cf. Eq.~\ref{eq:delta_self}). Hence, the value of $q$ which minimizes the free
energy is usually a result of the Cooper pair breaking processes occurring in several subbands and we can not distinguish between individual
contributions to the total momentum $q$ coming from each of them. From Fig.~\ref{fig3}(a) we can see that all of the FF phase stability regions are
extended over the magnetic field range in which the superconductivity is destroyed in several consecutive subbands. 
The widest one, starting with depairing in the subbands $(0, -4)$ and $(0, 2)$ extends up to $H=11.5$~T which is the critical field for the states:
$(0,-3)$, $(0,1)$, $(1,1)$ and $(1,2)$. The Cooper pair breaking occurring simultaneously in the four subbands is accompanied by the highest
jump in $\bar {\Delta}(H)$ which is largely caused by the fact that $(1,2)$ is the resonant state with the highest contribution to the superconducting
order parameter [see Fig.~\ref{fig5}(c)]. As shown in Fig.~\ref{fig4}(a), the formation of the FF phase for this particular case, requires to adjust
the Cooper pair momentum $q$ which is almost four times greater than that observed in other FF stability regions. Its high value is mainly determined
by the Fermi wave vector mismatch in the resonant subband $(1,2)$. The last FF phase stability region presented in Fig.~\ref{fig3}(a) is related
to the onset of depairing in the states $(0,-2)$ and $(0,0)$. Note that the orbital effect does not exists for states with $m=0$ and so the FF
phase related to depairing in the subband  $(0,0)$ is solely induced by the Zeeman effect. For this reason, the Cooper pair momentum $q$ in this
region is twice smaller than in the regions with the dominant role of the orbital effect [Fig.~\ref{fig4}(a)].

Calculations carried out for different values of $d$ show that the similar phase diagram, in which the FF phase stability regions are sandwiched between 
the standard BCS state stability ranges, is characteristic for each resonant thickness. As an example, in Fig.~\ref{fig3}(b) we present $\bar {\Delta} (H)$  for the neighboring
resonant point $d=1.19$~nm, for which the enhancement of the energy gap $\bar {\Delta}$ is due to the Cooper pairing in the state $(1,\pm 3)$ whose
bottom passes though the Fermi level. Interestingly, the FF phase is not formed for the non-resonant thickness, $d=1.39$~nm [Fig.~\ref{fig3}(c)], when
the superconductor to normal metal transition has a more BCS-like character without noticeable jumps in $\bar{\Delta}(H)$. For $d=1.39$~nm, the
spatially averaged value of the superconducting order parameter $\bar{\Delta}=0.142 \text{~meV}  < \Delta _{bulk}$ [Fig.~\ref{fig5}(f)]. All subbands
are far away from the Fermi level having almost equal contributions to the superconducting state. All this causes that deparing in an individual
subbands is less energy-consuming and consequently, the formation of the FF phase is unfavorable.

\begin{figure}[ht]
\begin{center}
\includegraphics[scale=1.0]{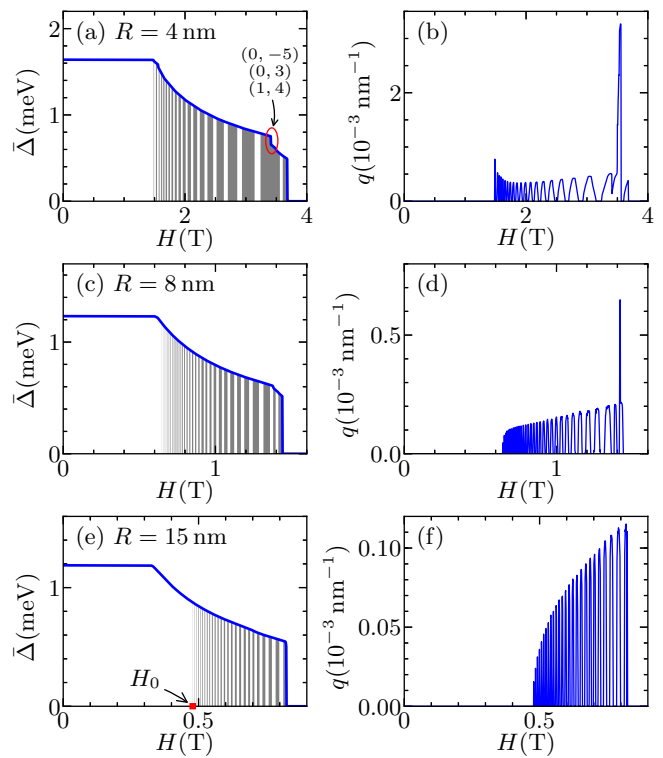}
\caption{(Color online) Magnetic field dependence of the averaged superconducting gap $\bar {\Delta}(H)$ (right panels) and total Cooper-pair momentum
$q(H)$ for the resonant shell thicknesses $d=1.09$~nm and nanowire radii (a,b) $R=4$~nm, (c,d) $R=8$~nm and (e,f) $R=15$~nm. In the right panels, gray
areas represent the FF states between which the conventional BCS phases, displayed as white regions, are stable.}
\label{fig6}
\end{center}
\end{figure}

\begin{figure}[ht]
\begin{center}
\includegraphics[scale=1.0]{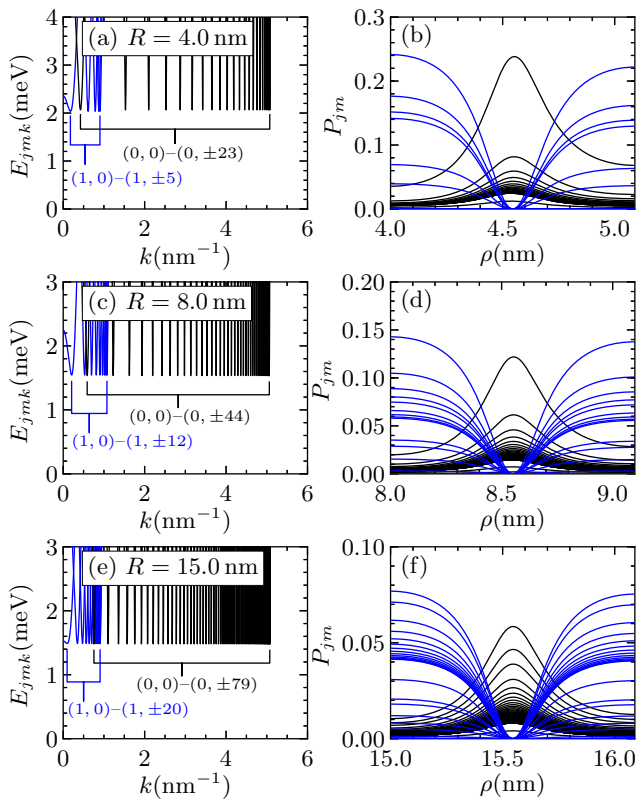}
\caption{(Color online) Quasiparticle dispersions $E_{kjm}$ vs $k$ (left panels), contributions to the paired phase coming from
different subbands $P_{j,m}$ (left panels) calculated for the resonant shell thickness $d=1.09$ and (a,b) $R=4$~nm, (c,d) $R=8$~nm, (e,f) $R=15$~nm.
Subbands with the radial quantum number $j=1$ are plotted by blue lines while black lines correspond to states with $j=0$. }
\label{fig7}
\end{center}
\end{figure}

Now, let us discuss how will the phase diagram change if we increase the nanowire radius $R$ up to the nanofilm-like regime, where $R /d \gg 1$.
In Fig.~\ref{fig6} (right panels) we present the magnetic field dependence of the spatially averaged superconducting gap $\bar {\Delta}(H)$ for the
resonant shell thickness $d=1.09$~nm (see Fig.~\ref{fig2}) and nanowire radii (a) $R=4$~nm, (c) $R=8$~nm and (e) $R=15$~nm. As previously,  the FF
phase stability ranges are displayed by gray areas while the corresponding values of the Cooper pair momentum $q$ are plotted in the left panels (b,d,f).
The phase diagrams in Fig.~\ref{fig6} differ considerably from that calculated for the nanowire-like regime, for $R=1$~nm (Fig.~\ref{fig3}).
The irregular pattern of the FF phase occurrence from Fig.~\ref{fig3}  evolves towards the regular distribution, in which the FF phases appear
periodically between the BCS state stability ranges. As discussed, with increasing nanowire radius, the centrifugal term of the single electron energy is
suppressed which, in turn, leads to the formation of bundles of subbands with the same radial quantum number $j$ and different $|m|$. Therefore, the
number of subbands $N_s$ taking part in the superconducting phase increases significantly. For $R=4$~nm, shown in Fig.~\ref{fig6}(a), $N_s=58$ and
the subbands $(0,0)-(0,\pm 23)$ and $(1,0)-(1,\pm 5)$ make a contribution to the paired state [see dispersion $E_{kjm}$ vs $k$ in
Fig.~\ref{fig7}(a)]. 
Such a large number of states $N_s$ makes the contribution of an individual subband to the superconducting order parameter less
significant. Consequently, the magnetically-induced depairing in a single subband is not so energy-consuming and it is not accompanied with the jump
in $\bar{\Delta}(H)$ as in the nanowire-like regime. Contrarily, as show in Fig.~\ref{fig6}, $\bar{\Delta}$ decreases rather smoothly with increasing
magnetic field up to $H_c$, at which the superconductor to normal metal transition is of the first order. Note that for $R=4$~nm we can still observe
the single small jump [marked by arrow in Fig.~\ref{fig6}(a)] which gradually disappears for larger $R$ and for $R=15$~nm it is not observed any
longer. Such residual jumps can occur in
the intermediate regime ($R/d \approx 1$), where a single subband contribution to the paired state can be still substantial (compare scales in the
right panels, in Fig.~\ref{fig7}). In this particular case, the jump is due to the simultaneous Cooper pair breaking in the subbands $(0,-5)$,
$(0,3)$, $(1,4)$ with the total contribution to the superconducting order parameter at $9$ \%.
As presented in Fig.~\ref{fig6}, the regions of the smooth decrease of $\bar{\Delta}(H)$ are divided into FF phases which appear
quasi-periodically alternating with the ordinary BCS paired state stability ranges. This periodicity is the more noticeable, the closer to the nanofilm-like regime
we approach - compare Fig.~\ref{fig6}(a) and (e). In the intermediate regime, for $R=4,8$~nm, the quasi-periodic pattern is disturbed
in the vicinity of the jump where the corresponding total momentum of the Cooper pair $q$ shows a distinct peak - see Fig.~\ref{fig6}(b). The reason
for this is the simultaneous Cooper pair breaking in three subbands which requires to adjust the Cooper pair momentum $q$ which is almost six
times larger than those obtained in the other FF stability regions. Similarly as the jump in $\bar{\Delta}(H)$, the peak in $q(H)$ disappears with
increasing $R$ and for $R=15$~nm it is not observed any longer. 

More detailed analysis of the FF phase formation in the nanofilm-like regime can be made based on Fig.~\ref{fig6}(e,f) for $R=15$~nm, where $R/d
\gg 1$. The regular occurrence of the FF phases presented in Fig.~\ref{fig6}(f) can be explained based on the same arguments as used in the
nanowire-like regime.
Namely, each of the FF stability regions is due to the Cooper pair breaking in the individual subbands while increasing magnetic
field. Since in the nanofilm-like regime $\Delta _{j,m}(H=0)$ do not depend on the quantum numbers (in contrary to the nanowire-like
regime), $H_c^{j,m}=H_c^{j,-m-2}$ and $H_c^{j,m_1}>H_c^{j,m_2}$ for any two states with positive $m_1<m_2$. It means that the Cooper pair
breaking starts from the states $(0,M)$ and $(0,-M-2)$, where $M$ is the highest positive orbital magnetic quantum number and subsequently, it
takes place in the subbands $(j,m)$ and $(j,-m-2)$ with $m=M,M-1,\ldots,0$. It is of interest that, regardless of the number of states $N_s$, the
first FF phase region is derived by the Cooper pair breaking in the subbands $(0,38)$ and $(0,-40)$. The FF phase corresponding to depairing in
the states with higher $|m|$ do not occur. As an example, for $R=15$~nm, $N_s=200$, the subbands $(0,0)-(0,\pm79)$ and $(1,0)-(1,\pm20)$ make a
contribution to the paring state [see Fig.~\ref{fig7}(e)] and although depairing in the states with high $|m|$ starts at $H\approx 0.34$~T, at
which $\bar{\Delta}$ starts decreasing [see Fig.~\ref{fig6}(e)], the first FF phase occurs for $H\approx 0.47$~T where the Cooper pair breaking takes
place in the subbands $(0,38)$ and $(0,-40)$. It explains the gradual shift of the region where the FF phase stability regions occur, towards $H_c$ for larger $R$.
We expect that in the limit $R/d \rightarrow \infty$ this region moves to the close vicinity of $H_c$ and, due to the induced degeneration with
respect to $m$, all FF phase stability regions will merge into one. This picture is consistent with the ordinary FF phase diagram predicted for
nanofilms.\cite{Matsuda2007} 
In Fig.~\ref{fig6}(f) we can also observe that the subsequent FF phase regions become narrower with decreasing magnetic field up to value
at which the Cooper pair breaking occurs in the subbands $(0,38)$ and $(0,-40)$. Below this critical value $H_0$, the FF phases do not occur.
Simultaneously,  the corresponding value of $q$ tends to zero for $H=H_0$ [see Fig.~\ref{fig6}(f)]. This characteristic behavior can be explained as
resulting from the difference in the orbital energy acquired from the magnetic field by the states with different orbital magnetic quantum number.
Since the orbital term $m \mu _B H$ is proportional to the quantum number $m$, in the presence of the magnetic field, the states with higher $|m|$
acquire the orbital energy much greater than the states with lower $|m|$. As a result the Fermi wave vector mismatch $q_{j,m}$ between the paired
electrons $(k,j,m,\uparrow)\leftrightarrow (-k,j,-m,\downarrow)$ is larger for states with higher $|m|$. If the Cooper pairs are broken in the
state with high $|m|$, in a certain magnetic field, the wave vector mismatch $q_{jm}$ in the states with low $|m|$ is still very small. Since the
value of $q$ is a result of the Fermi vector mismatches $q_{j,m}$ in all states contributing to the superconductivity, the formation of phase with a
nonzero $q$ is energetically unfavorable. 
The critical is depairing in the subbands $(0,38)$ and $(0,-40)$, when the Fermi vector mismatches in all superconducting states become
sufficiently large to gain the small value of $q$ by all these subbands. The FF phase with such a small value of $q$ is
very susceptible to the magnetic field and even slight increase of $H$ causes that this phase is destroyed and the system switches back to the BCS
paring - note that the first FF phase stability region corresponding to depairing in the states $(0,38)$ and $(0,-40)$ is extremely narrow. The same
behavior is repeated each time when the Cooper pairs are broken in subsequent subbands while increasing magnetic field. However, for the states
with lower $|m|$, the Fermi wave vector mismatches become larger. Consequently, the reduction of the so-called depairing region on the Fermi
sphere requires larger momentum vector $q$, as shown in Fig.~\ref{fig6}(f). Since larger $q$ requires a higher magnetic field needed to destroy the FF
phase, we observe the gradual extension of the FF phase stability regions for higher magnetic field. 

\section{Conclusions and outlook}
\label{sec:conclusion}
The orbital effect on the FF phase has been investigated in superconducting core/shell nanowires subjected to the axial magnetic
field. The energy quantization induced by the confinement of the electron motion in the radial direction leads to the multiband superconductivity,
similarly as found in novel superconductors, e.g., MgB$_2$ or iron pnictides. It reveals in the form of the quantum size oscillations,
i.e. the spatially averaged energy gap $\bar {\Delta}$ varies with $d$ at a fixed $R$. The character of $\bar {\Delta}$ variations changes
considerably with increasing $R$. From irregular pattern typical for nanowires, it evolves to the regular oscillations characteristic for nanofilms,
while the crossover between both the regimes is determined by the centrifugal energy. As discussed above, in superconducting core/shell
nanowires, the orbital effect  which so far has been considered as detrimental to the FF phase formation, can in fact induce the non-zero momentum
paired state. In the presence of magnetic field, the degeneracy with respect to the orbital magnetic quantum number $m$ is lifted which leads to
the Fermi wave vector mismatch between the subbands with opposite orbital momenta in the paired state. Therefore, as the magnetic field increases, the
superconductivity is destroyed in subsequent subbands which manifests itself as a cascade of jumps in $\bar{\Delta}(H)$. To sustain the Cooper
pairing $(k,j,m,\uparrow)\leftrightarrow (-k,j,-m,\downarrow)$ in the corresponding subband, the non-zero total momentum state (FF
phase) is formed which, in turn, leads to the phase diagram of alternating FF and BCS stability regions. 
In the  present paper, by controlling of the ratio $R/d$,  we have switched from the nanowire-like ($R/d \ll 1$) to the nanofilm-like ($R/d \gg 1$)
scenario, strengthening or suppressing the centrifugal energy, respectively.
We have found that the phase diagrams differ considerably in both regimes. The irregular pattern of the FF phase occurrence in the nanowire-like
regime evolves towards the regular distribution, in which the FF phase stability regions appear periodically between the BCS state stability regions, 
in the nanofilm-like regime. As
presented, the crossover between these two different phase diagrams can be explained as resulting from the orbital effect and the multigap character
of superconductivity. In the nanowire-like regime, the centrifugal term, $\hbar ^2 m^2/2m_er^2$, contributes significantly to the single electron
energy, which leads to the well separated states for different $|m|$. Due to the quantum confinement and Andreev mechanism, induced by the spatially
dependent superconducting order parameter, the system exhibits multiband and multigap superconductivity, in which $\Delta _{j,m}$ may vary
for different quantum numbers. In the presence of magnetic field, the orbital effect leads to a situation in which the Fermi wave vector mismatch
between the paired electrons $(k,j,m,\uparrow)\leftrightarrow (-k,j,-m,\downarrow)$ varies with $m$. The critical magnetic field for the 
subband $(j,m)$ is given by $H_c^{j,m}=\Delta _{jm}/(|m+1|)\mu_B$. As shown, the Cooper pair breaking in each of the subbands entails the formation of
the FF phase. Therefore, due to the multigap character of superconductivity and  irregular position of states with
different $m$ on the energy scale, the FF phase occurrence shows irregular pattern in the nanowire-like regime. This picture changes considerably if
we increase $R$ up to the limit $R/d \gg 1$.  In the nanofilm-like regime the centrifugal term is negligibly small and the states with different $|m|$
for $H=0$ are almost degenerate forming the bands labeled by the radial quantum number $j$. The multigap character of the
superconductivity vanishes, i.e. all subbands $(j,m)$ have the same value of $\Delta _{jm}$. Consequently, FF phases start to occur
quasi-periodically each of which is related to the Cooper pair breaking in subsequent subbands with decreasing $|m|=M,M-1,...$. In the limit $R/d
\rightarrow \infty$, our explanation leads to the phase diagram consistent with that predicted for nanofilms.

Although  the presented transitions between FF and BCS phases seem to be verifiable by the use of standard experimental techniques, e.g., the
specific heat measurements or detection of a supercurrent induced by the magnetic field, the FF phase appearance in the
realistic core/shell nanowires requires some additional remarks, especially with regard to the assumptions made in the theoretical model. 
First, for the sake of simplicity, in this work we consider the Fulde-Ferrell phase which assumes the single Cooper pair momentum $q$ for all
subbands contributing to the superconducting state. Since, due to the orbital effect, the magnetically-induced Fermi wave vector mismatch is
different for different subbands, it would be interesting to study the case when the superconducting order parameter is a combination of many
components with different $q$ vectors. In this manner, we would be able to reduce the so-called  depairing region on the Fermi sphere to a greater
extent what, in effect, would minimize the free energy of the system to the value much lower than that obtained for the FF phase. In
fact, this energetically more favorable multi-momentum Cooper pair state could be observed in experiment but its appearance does not alter
the qualitative picture of the non-zero momentum phase creation presented in the paper. The serious limitation in the experimental studies of the
FF phase in metallic nanostructures is the appearance of impurities as the FF state could be readily destroyed by scattering. Even in ultra-clean nanowires, the
surface scattering can destroy the FF state. However, as discussed above, the phase diagram of alternating FF and BCS stability regions emerges only for
the resonant thicknesses when the superconducting order parameter is controlled by the  single-electron subbands whose bottom is
situated in the vicinity of the Fermi level. Its characteristic location on the energy scale causes that the corresponding longitudinal wave vector
$k<0.5$~nm$^{-1}$. It corresponds to the electron wave length grater than $50$~nm. Propagation of such long waves should be insensitive to the local
surface imperfection with the size of the unit cell. This argument is even stronger if we consider the wave length corresponding to the FF phase for
which the total momentum of the Cooper pairs $q<0.01$~nm$^{-1}$ gives the wave length greater than $600$~nm.

Finally, we would also like to address fluctuations which may appear in superconducting low dimensional structures. Thermally
activated phase slip and quantum phase slip are known to play a serious role in superconducting nanowires making the use of the mean field theory
questionable.
However, as shown by recent experimental studies of superconducting Pb nanofilms, the use of the mean-field theory seems to be justified giving
a surprisingly good agreement with experiment for the nanofilm thickness down to $2-5$ monolayers.\cite{Qin2009} For superconducting nanowires, the
predicted diameter limit is $5-8$~nm,\cite{Arutyunov2008} below which the quantum-phase slip can suppress the superconductivity. However, recent
experiments for nanowires with diameter $5-6$~nm\cite{Altomare2006} do not show any signature of phase fluctuations. Thus,
we may expect that the mean-field approach used in the paper is reasonable for considered geometry of the core/shell nanowires for which
phase fluctuations are assumed not to occur. 

\section{Acknowlegement}
This research was supported in part by PL-Grid Infrastructure. The authors acknowledge valuable discussions with M. Zegrodnik.

%

\end{document}